\documentclass[a4paper]{jpconf}
\usepackage{epsfig} 
\usepackage{amsmath}
\usepackage{placeins}
\usepackage[pass,a4paper]{geometry}
\usepackage{epstopdf}

\begin{document}
%\linenumbers
\title{Light flavour hadron production in pp collisions at $\sqrt{s} = $ 13 TeV with ALICE}
\author{Ra\'ul Tonatiuh Jim\'enez Bustamante for the ALICE Collaboration}
\address{GSI Helmholtzzentrum f{\"u}r Schwerionenforschung GmbH,
Planckstra{\ss}e 1, 64291 Darmstadt, Germany  and Ruprecht-Karls Heidelberg University, Seminarstra{\ss}e 2, 69117 Heidelberg, Germany.\\}
\ead{raul@cern.ch}

\begin{abstract} 

The ALICE detector has excellent Particle IDentification (PID) capabilities in the central barrel ($\lvert \eta \rvert <$ 0.9). This allows identified hadron production to be measured over a wide transverse momentum ($p_{\rm{T}}$) range, using different sub-detectors and techniques: their specific energy loss (d$E$/d$x$), the velocity determination via time-of-flight measurement, the Cherenkov angle or their characteristic weak decay topology. Results on identified light flavour hadron production at mid-rapidity measured by ALICE in proton-proton collisions at $\sqrt{s}$ = 13 TeV are presented and compared with previous measurements performed at lower energies. The results cover a wide range of particle species including long-lived hadrons, resonances and multi-strange baryons over the $p_{\rm{T}}$ range from 150 MeV/\emph{c} up to 20 GeV/\emph{c}, depending on the particle species.

\end{abstract}

\section{Introduction}

In June 2015, the Large Hadron Collider (LHC) has resumed its physics program with proton-proton (pp) collisions at $\sqrt{s}$ = 13 TeV. Currently, this is the highest centre-of-mass energy reached in the laboratory. A full description of particle production in pp collisions cannot be obtained from perturbative Quantum Chromodynamics and hence modelling efforts typically employ various empirical components that have to be adjusted based on experimental data. The measurement of light flavour particles production in pp collisions provides important input for event generators to model the soft parton interactions and the hadronization processes. Furthermore, pp collisions are used as reference to isolate medium effects present in larger colliding systems. Previous ALICE measurements \cite{pp900gev}, \cite{pp276tev}, \cite{pp7tev} on identified particles in pp collisions at lower energies show that the models usually give a fair description of the shapes of the $p_{\rm{T}}$ spectra but fail on the description of the pion, kaon and proton yields. 
%Some ideas exploring the differences between the underlying mechanisms in the different event generators in order to understand the particle production in pp collisions have been proposed in \cite{radialflow}, \cite{sphericity}.

\section{Particle Identification}

$\pi^{\pm}$, K$^{\pm}$, p and $\bar{\mbox{p}}$ are identified in the rapidity window $\lvert  y \rvert  <$ 0.5  using several Particle IDentification (PID) techniques in the different ALICE sub-detectors: the Inner Tracking System (ITS), the Time Projection Chamber (TPC), the Time Of Flight detector (TOF) and High Momentum Particle Identification Detector (HMPID). The combined information of these detectors in different $p_{\rm{T}}$ regions allows us to measure the particle production in the region starting from 150 MeV/\emph{c} up to 20 GeV/\emph{c}. 

The $K^{*0}$ and $\phi$ mesons are measured with invariant-mass analyses via the reconstruction of their hadronic decays. The combinatorial background is subtracted using either a like-sign or event mixing technique. The resulting invariant-mass distributions are then fitted. Previous measurements by ALICE and details on the analysis procedure can be found in \cite{ppresonances}.

The multi-strange $\Omega^{+}$, $\Omega^{-}$, $\Xi^{+}$, $\Xi^{-}$ baryons are reconstructed via their weak hadronic decay products. Candidates are selected with restrictions on the topology of the decay and the signal is extracted from the resulting invariant mass distributions. The full analysis procedure is described in \cite{hyperons}. 

\section{Results}

\begin{figure}[h!] 
\begin{center}
\epsfig{file=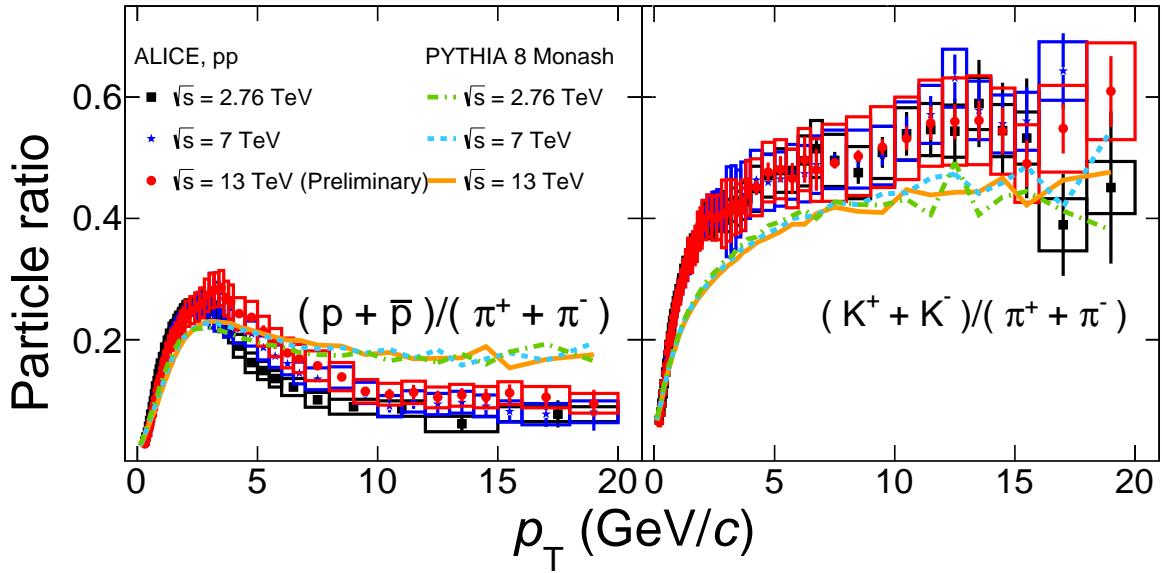, bb=0 0 567 280,clip=,angle=0,width=1.0\textwidth}
 \caption {Ratios ($p + \bar{p})/ (\pi^{+} + \pi^{-})$ and $(K^{+} + K^{-})/ (\pi^{+} + \pi^{-})$ in pp collisions at $\sqrt{s}$ = 2.76, 7.0 and 13 TeV. The lines show PYTHIA 8 Tune Monash-2013 \cite{pythiamonash} predictions for the 3 different energies.}
\label{ratios}
\end{center}
\end{figure}

The $p_{\rm{T}}$-dependent $(p + \bar{p})/ (\pi^{+} + \pi^{-})$ and $(K^{+} + K^{-})/ (\pi^{+} + \pi^{-})$ ratios measured in central rapidity ($ |y| <$ 0.5) at $\sqrt{s}$ = 2.76, 7.0 and 13 TeV are shown in Fig. \ref{ratios}. The comparison to PYTHIA 8 \cite{pythia8} (Monash-2013 \cite{pythiamonash}) for the different energies is also shown.

Within the systematic uncertainties the $(K^{+} + K^{-})/ (\pi^{+} + \pi^{-})$ ratio is consistent for the three different energies. The $(p + \bar{p})/ (\pi^{+} + \pi^{-})$ ratio for the low-intermediate $p_{\rm{T}}$ region shows a slight shift towards higher $p_{\rm{T}}$ when increasing $\sqrt{s}$. Within our systematic uncertainties the ratio is compatible for the three different energies in the $p_{\rm{T}}$ region above 10 GeV/\emph{c}. This shift is also observed on the Monte Carlo predictions from the PYTHIA 8 model. The origin of the peak in PYTHIA 8 is attributed to the colour reconnection mechanism \cite{colorreconnection}, and is expected to become more important at higher centre-of-mass energy.

%Previous ALICE measurements at $\sqrt{s}$ = 2.76 TeV and $\sqrt{s}$ = 7 TeV  are reported in \cite{pidalicepapers} are also shown.

\begin{figure}[h!] 
\begin{center}
\minipage{0.45\textwidth}
\epsfig{file=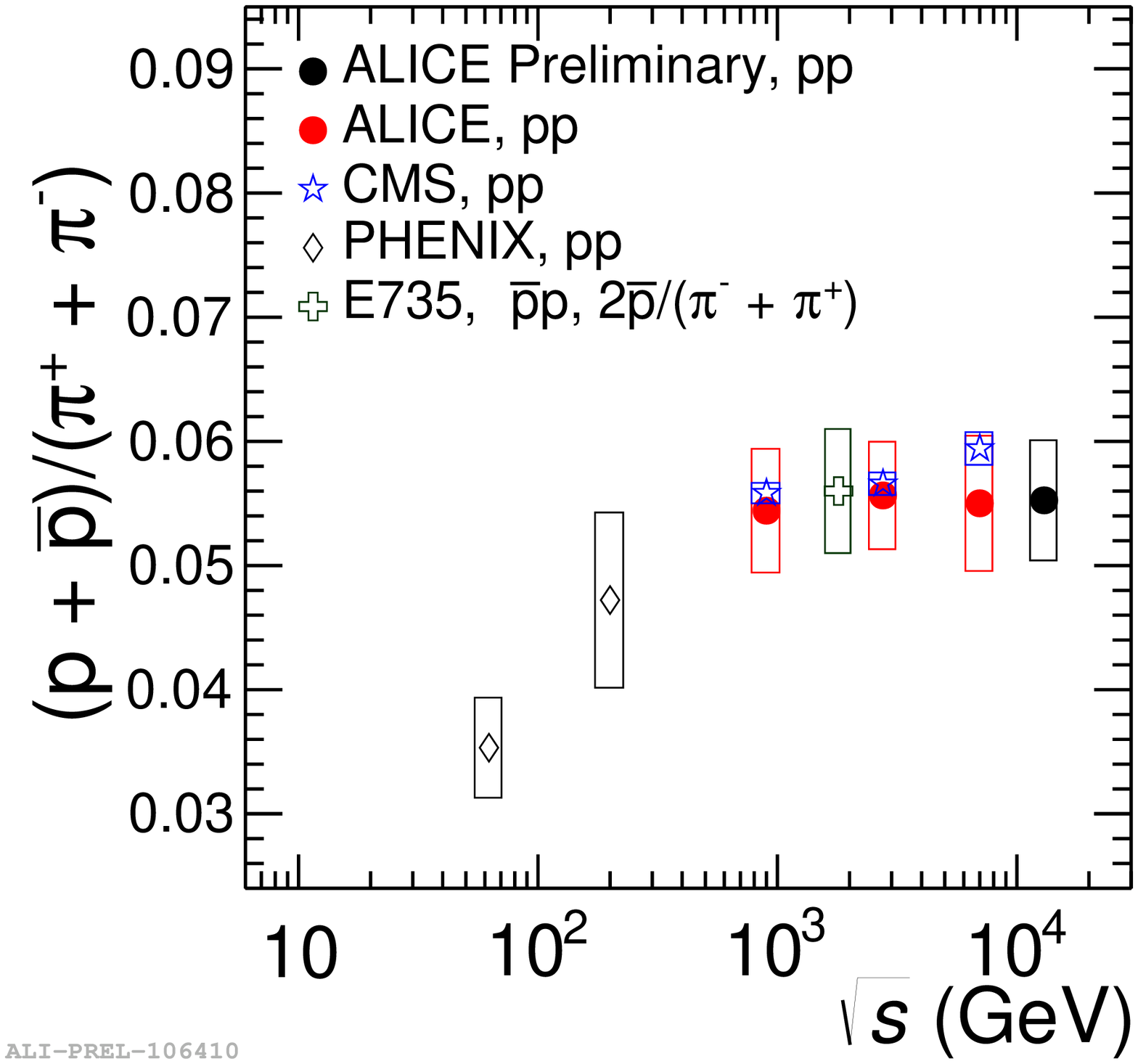, bb=0 0 567 541,clip=,angle=0,width=1.0\textwidth}
 \endminipage   
 \hspace{5px}
 \minipage{0.45\textwidth}
 \epsfig{file=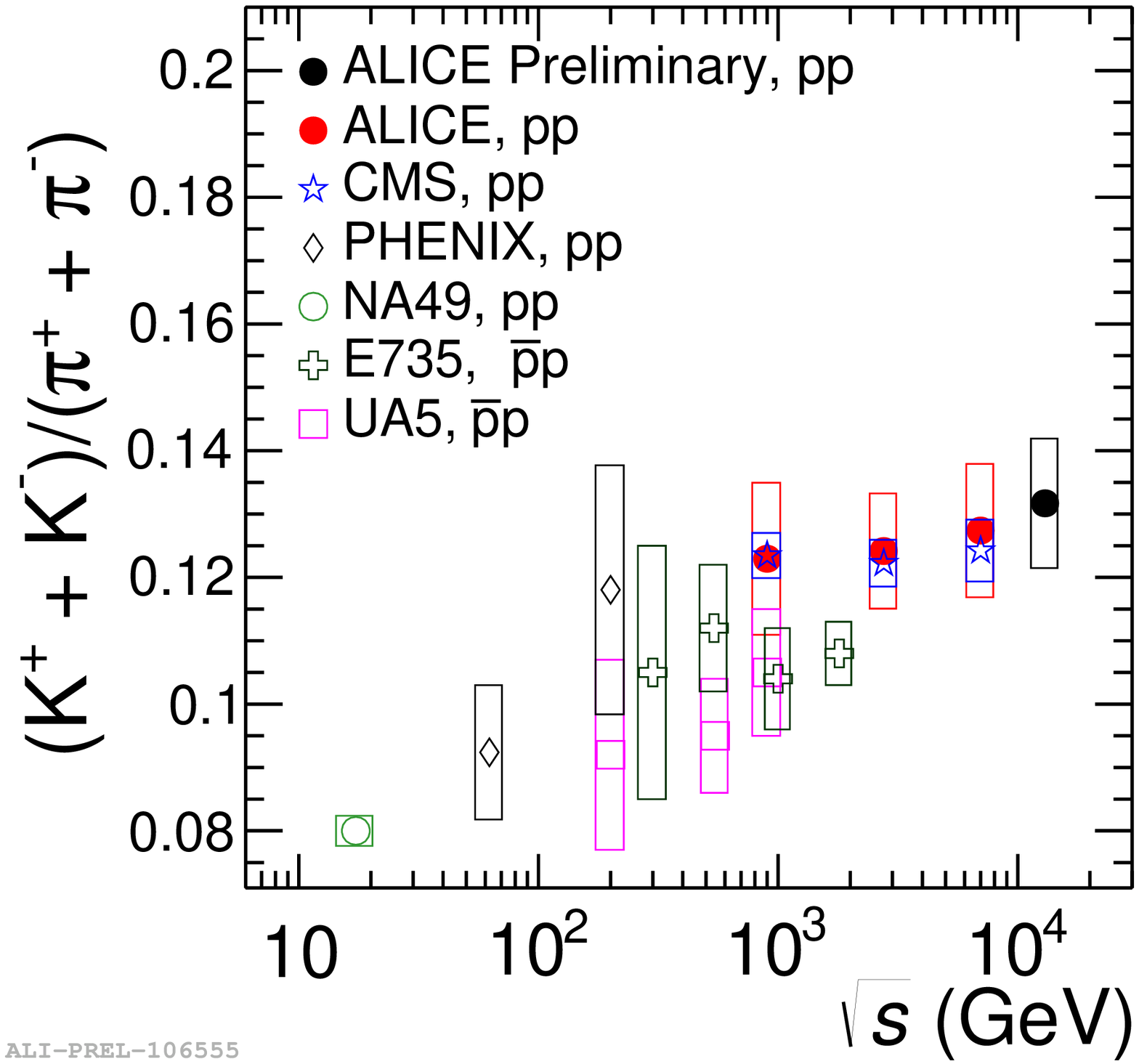, bb=0 0 567 541,clip=,angle=0,width=1.0\textwidth}
 \endminipage
 \caption { $(p + \bar{p})/ (\pi^{+} + \pi^{-})$ and $(K^{+} + K^{-})/ (\pi^{+} + \pi^{-})$ in pp  and p$\bar{\mbox{p}}$ collisions as a function of $\sqrt{s}$. Errors bars represent the quadratic sum of statistical and systematics uncertainties.} 
\label{ratios_energies}
\end{center}
\end{figure}

The $p_{\rm{T}}$-integrated  $(K^{+} + K^{-})/ (\pi^{+} + \pi^{-})$ and  $(p + \bar{p})/ (\pi^{+} + \pi^{-})$ ratios in pp collisions as a function of $\sqrt{s}$ are shown in Fig.\ref{ratios_energies}. Results at different energies for different experiments are also shown. Results from E735 at $\sqrt{s}$ = 0.3, 0.54, 1.0 and 1.8 TeV \cite{resultse735}, \cite{resultse7352}, from PHENIX at $\sqrt{s}$ = 62.4 and 200 GeV \cite{resultsphenix} and CMS at  $\sqrt{s}$ = 0.9, 2.76 and 7 TeV \cite{resultscms} are compared for both particle ratios. The last added point at 13 TeV confirms the similar trend and saturation for $\sqrt{s}>$ 900 GeV observed in ALICE previous measurements.

\begin{figure}[h!] 
\begin{center}
\epsfig{file=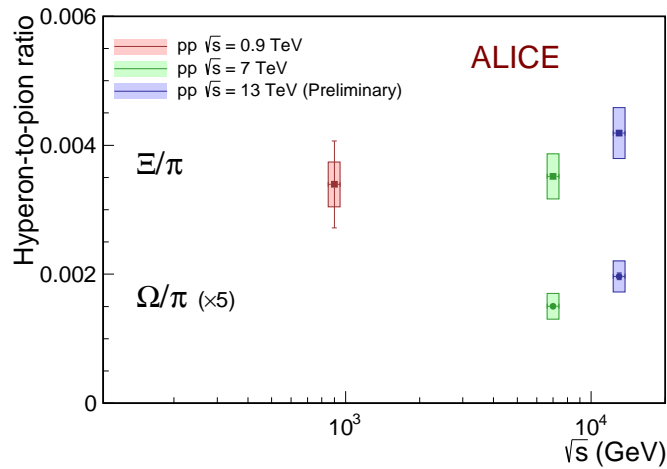, bb=0 0 567 408,clip=,angle=0,width=0.55\textwidth}
 \caption {Hyperon to pion ratio in pp collisions as a function of the collision energy.}
\label{hyperon_ratio}
\end{center}
\end{figure}

The hyperon-to-pion ratio as a function of $\sqrt{s}$ measured in ALICE is shown in Fig. \ref{hyperon_ratio}. In comparison with the kaon- and proton-to-pion ratio the  strange- and multi-strange-to-pion ratios show a hint of increase at 13 TeV with respect to the lower collision energies. This enhanced production may be related to the enhancement observed as a function of multiplicity in pp collisions at $\sqrt{s}$ = 7 TeV \cite{multistrangeppmult}, as the charged particle multiplicity density increases with the collision energy.

\begin{figure}[h!] 
\begin{center}
\minipage{0.45\textwidth}
\epsfig{file=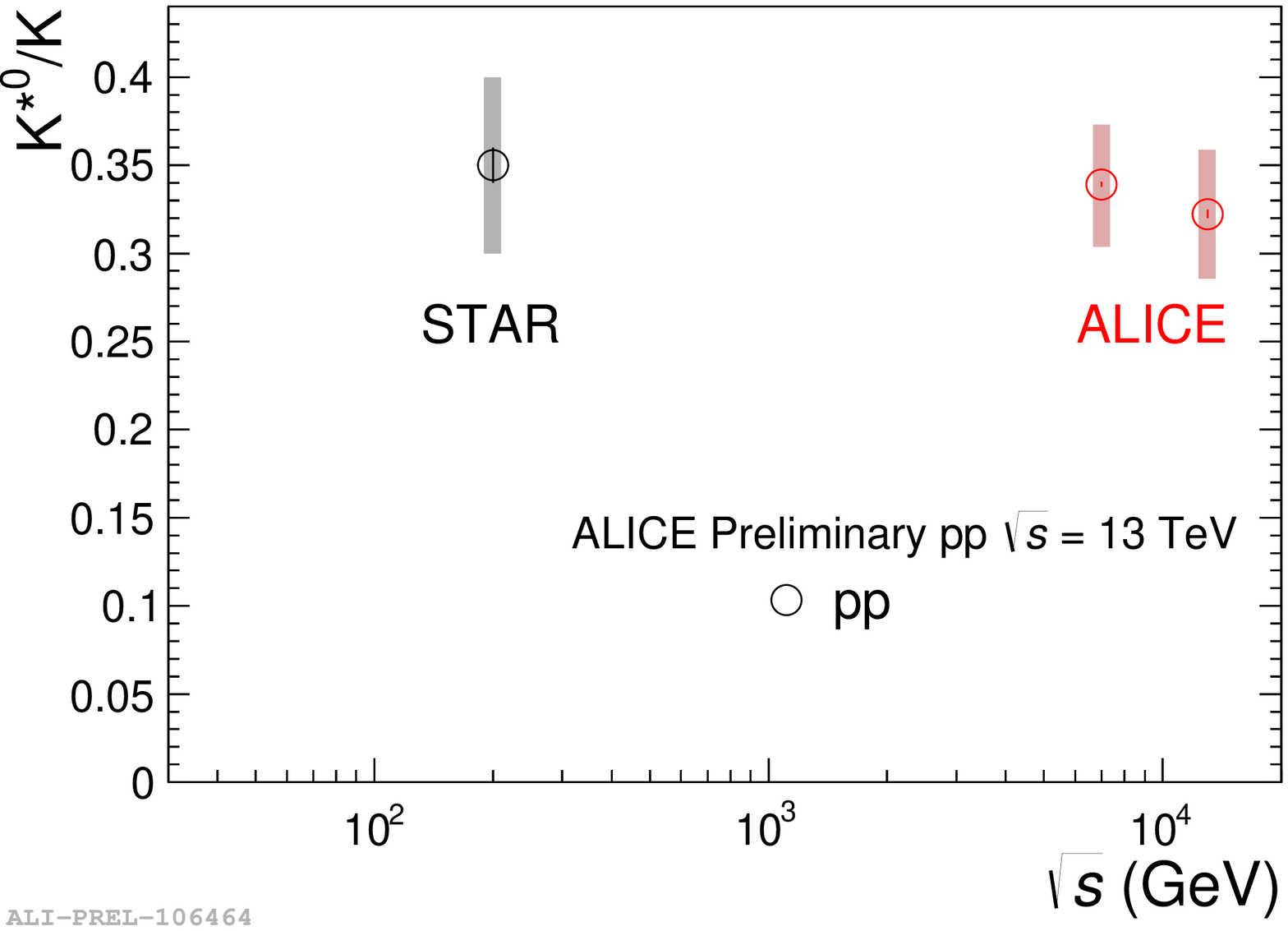, bb=0 0 567 409,clip=,angle=0,width=1.0\textwidth}
 \endminipage   
 \hspace{5px}
 \minipage{0.45\textwidth}
 \epsfig{file=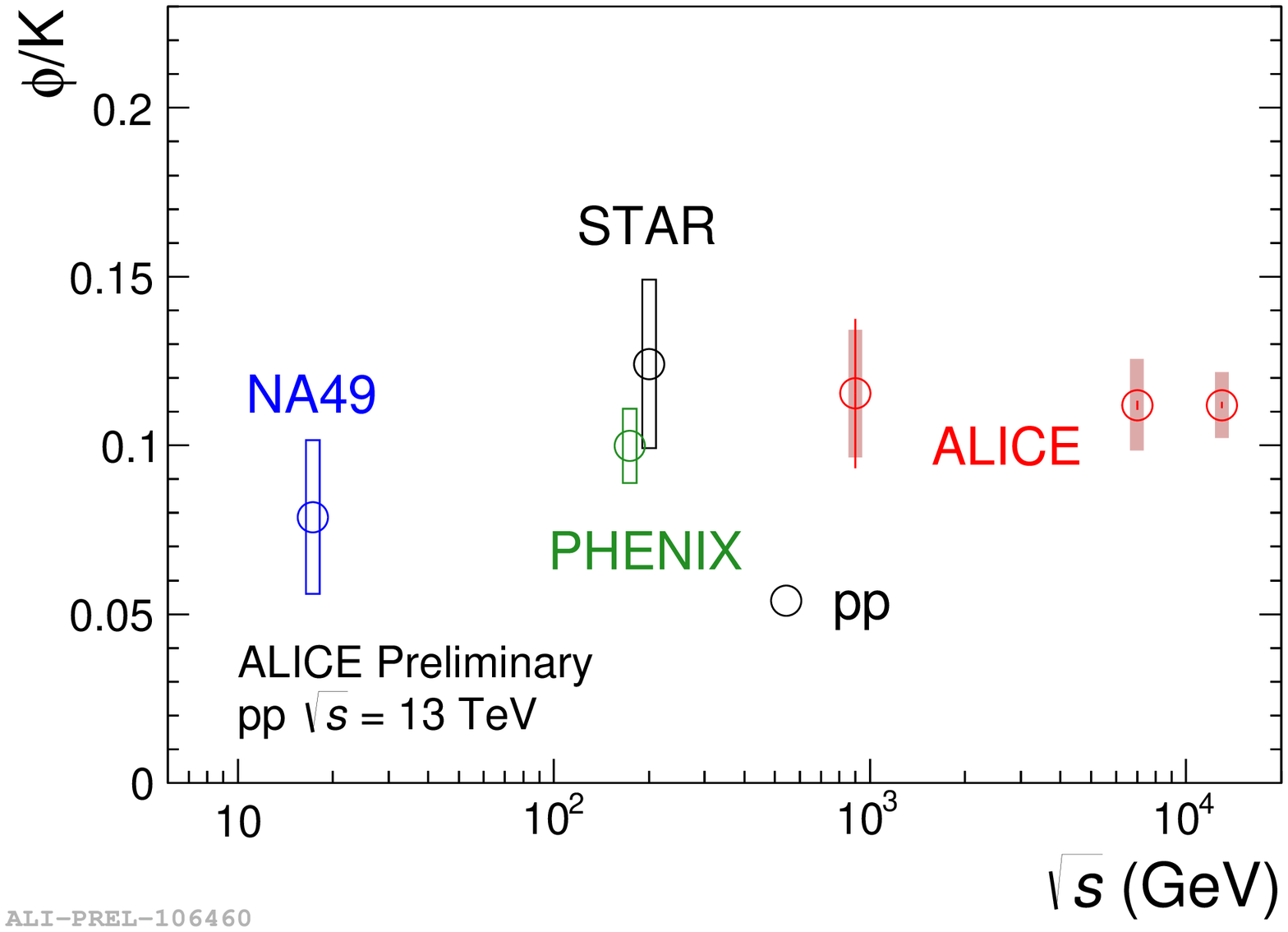, bb=0 0 567 409,clip=,angle=0,width=1.0\textwidth}
 \endminipage
 \caption { $K^{*0}/K$ and $\phi/K$ ratio in pp collisions as a function of the collision energy. Bars represent statistical uncertainties. Boxes represent the total systematic uncertainties.} \label{ratios_resonances}
\end{center}
\end{figure}

The $K^{*0}$ and $\phi$ meson production has been also measured in pp collisions at $\sqrt{s}$ = 13 TeV. The $p_{\rm{T}}$-integrated $K^{*0}/K$ and $\phi/K$ ratios are shown in Fig. \ref{ratios_resonances}. However, no energy evolution is observed in the range from RHIC to the highest LHC energy. 

\section{Conclusions}
The $p_{\rm{T}}$ dependence of the $(p + \bar{p} )/ (\pi^{+} + \pi^{-})$ and $(K^{+} + K^{-})/ (\pi^{+} + \pi^{-})$ is the same at LHC energies within our systematic errors. PYTHIA 8 describes the shape but exhibits large deviations from the data for transverse momenta larger than 2 GeV/\emph{c}. 
Above $\sqrt{s} = $ 900 GeV the $(p + \bar{p} )/ (\pi^{+} + \pi^{-})$ and $(K^{+} + K^{-})/ (\pi^{+} + \pi^{-})$ ratios are energy independent within systematic uncertainties. The relative production of strange resonances ($K^{*0}/K$ and $\phi/K$) remains constant within a wide range of centre of mass energy.
Multiplicity dependent studies in pp collisions at 13 TeV will help disentangle the role of collision energy and event multiplicity in the identified particle production of pp collisions.

\end{document}